\documentstyle{article}
\def\PL #1 #2 #3 {{\rm Phys. Lett.} {\bf#1} (#3) #2}
\def\NP #1 #2 #3 {{\rm Nucl. Phys.} {\bf#1} (#3) #2}
\def\ZP #1 #2 #3 {{\rm Z. Phys.} {\bf#1} (#3) #2}
\def\PRL #1 #2 #3 {{\rm Phys. Rev. Lett.} {\bf #1} (#3) #2}
\def\PR #1 #2 #3 {{\rm Phys. Rev.} {\bf#1} (#3) #2}
\def\MPL #1 #2 #3 {{\rm Mod. Phys. Lett.} {\bf#1} (#3) #2}
\def\RMP #1 #2 #3 {{\rm Rev.~Mod. Phys.} {\bf#1} (#3) #2}
\def\ifm{\ifmmode}

\def\als{\ifm \alpha_s \else $\alpha_s$\fi}
\def\go{\ifm \rightarrow \else $\rightarrow$\fi}

\def\eps{\ifm \epsilon \else $\epsilon $\fi}

\newcommand{\notp}{\ \hbox{{$p$}\kern-.43em\hbox{/}}}
\newcommand{\notE}{\ \hbox{{$E$}\kern-.43em\hbox{/}}}
\newcommand{\beq}{\begin{equation}}
\newcommand{\eeq}{\end{equation}}
\newcommand{\beqn}{\begin{eqnarray}}
\newcommand{\eeqn}{\end{eqnarray}}
\newcommand{\beqs}{\begin{eqnarray*}}
\newcommand{\eeqs}{\end{eqnarray*}}

\def\smin{s_{\rm min}}

\def\ts2p3{s_{2'3}}

\def\be{\begin{equation}}
\def\ee{\end{equation}}
\def\bea{\begin{eqnarray}}
\def\eea{\end{eqnarray}}

\begin{document}
\thispagestyle{empty}
\begin{flushright}
\hfill{CERN-TH/96-230}\\
\hfill{FERMILAB-CONF-96/306-T}
\end{flushright}

\vskip 2cm
\begin{center}
HIGHER ORDER QCD CORRECTIONS TO TAGGED PRODUCTION PROCESSES~\footnote{
Talk given by S. K. at the DPF96 Conference, Minneapolis, 
MN, August~10--15, 1996, to appear in the Proceedings
}
\vglue 1.4cm
\begin{sc}
Stephane Keller\\
\vglue 0.2cm
\end{sc}
{\it Fermilab, MS 106\\
Batavia, IL 60510, USA}
\vglue 0.5cm
and
\vglue 0.5cm
\begin{sc}
 Eric Laenen\\
\vglue 0.2cm
\end{sc}
{\it CERN TH-Division\\
1211-CH, Geneva 23, Switzerland}
\end{center}
 
\vglue 1.5cm
\begin{abstract}
\par \vskip .1in \noindent
We extend the phase space slicing method to allow for heavy quarks and 
fragmentation functions.  The method can be used to calculate 
differential cross section
in which any particular particle (massive or massless) is tagged.
\end{abstract}
\vfill
\begin{flushleft}
CERN-TH/96-230\\
FERMILAB-CONF-96/306-T
\end{flushleft}

\newpage

A next-to-leading order calculation has well known advantages 
compared to a leading order calculation:
it reduces the normalization uncertainty,
it starts to reconstruct the parton shower, and
it tests the convergence of the QCD perturbative expansion. 
Over the last few years the emphasis has been on constructing 
Monte-Carlo programs that include all $\als$ 
corrections and are fully differential in the final state particle momenta, 
such that any experimental cuts can be imposed.
One such method is the so-called 
``phase space slicing'' method.
The formalism for this method developed by Giele, 
Glover and Kosower~\cite{GGK} 
introduced a high degree of automation 
by using the following ingredients: decomposition of the amplitude 
according to the color structure into colorless subamplitudes, 
factorization of the phase space 
and of the colorless subamplitudes in the soft and collinear region,
and the generalization of crossing to NLO.
We have extended this particular formalism to allow for 
heavy quarks and fragmentation functions.  In this short contribution, 
we briefly review the different aspects of the
method, full details will be presented elsewhere~\cite{KL96}.  
The method has been used in Ref.~\ref{GKL} to calculate the
QCD corrections to $W+$ heavy quark production at the Tevatron.

The QCD corrections consist of virtual and real corrections.
The virtual corrections are the interference between the LO and 
all the one loop diagrams and  must be calculated in $n=4+\eps$ 
dimension in 
order to regularize the singularities.
Coupling constant and mass renormalization take care of the ultraviolet 
singularities, and require the introduction of the renormalization scale.
At the end, some collinear and soft singularities remain as 
$1/\eps$ and $1/\eps^2$ poles.    
The real corrections are those contributions with one more parton than the 
LO and have soft and/or collinear singularities.  
One begins by considering the processes were all the quarks and gluons 
are in the final state, e.g., 
$V \go q \bar{q} + n$ gluons, where
V stands for an electroweak gauge boson, such that all the 
singularities cancel without having to do mass factorization.  
The basic idea of the phase space slicing method 
is to separate the phase space in two regions using the 
invariants $S_{ij}=2\, P_i.P_j$, where the $P_i$ are 
the momentum of the final 
state particles.  The hard region is defined so that 
all the $S_{ij}$ are bigger than a theoretical cut-off $S_{min}$.  
In this case, the calculation can be done numerically in $n=4$ dimensions.  
The collinear and soft 
region is defined such that one or two $S_{ij}$ are smaller than 
$S_{min}$.  In this case, the calculation must be done analytically 
in $n=4+\eps$ dimensions.  
If $S_{min}$ is small enough, the soft and collinear approximation can be used
such that the integration in $n$ dimensions is greatly simplify.  
We have generalized the soft approximation to the case where the 
particles involved are massive.  In the collinear region, the mass regularizes 
the singularities and the calculation can be done numerically.  
The $1/\eps$ and $1/\eps^2$ poles that remain after the integration over 
the soft and collinear region cancel with 
the corresponding poles of the virtual contributions.  
At the end we are left with a ``K'' factor, proportional 
to the born cross section and dependent on $S_{min}$, the finite part 
of the virtual contribution, and the real corrections in the hard region, 
that also depends on $S_{min}$.  An important numerical test is that any 
observables should be independent of $S_{min}$.

The generalization of crossing to NLO is done through the use of the 
so-called ``crossing functions''.  Let us consider the process:
$p p \go V + n$ jets.   
The basic idea here is that we do not want to redo the cancellation of 
all the poles, but rather use the ``K'' factor already derived for 
$V \go (n+2)$ jets.  
First, the usual crossing for all the matrix elements
is done, along with the crossing of the finite terms of the virtual piece 
and the crossing of the ``K'' factor with appropriate analytical continuation.
Then two corrections must be applied in $n=4+\eps$ dimensions: 
1) subtraction of some collinear singularities included in ``K'' that are 
not present when
the born is crossed, 2) add some initial state collinear singularities 
not yet included in ``K''.  Along with mass factorization, all these corrections 
give terms that are proportional to the Born cross section: 
\beq
\als \, \sum_{a,b}\int dx_1\int dx_2
\,  C_a^{H_1,scheme}(x_1) \, {f}_b^{H_2}(x_2) \, 
d\sigma_{ab}^{LO}(x_1,x_2)\ .
\eeq
$f_a^H$ is the distribution function of parton a inside of the hadron H,
and the $C_a^H$ are the factorization scheme 
dependent crossing functions:  
\beq
C_a^{H,scheme}(x,\mu_F^2) = A_a^H(x)\, \ln(\frac{\smin}{\mu_F^2})
+B_a^{H,scheme}(x),
\label{crossing}
\eeq
where $\mu_F$ is the factorization scale.
The crossing function are universal and 
only need to be calculated once for a given set of parton distribution 
functions.  $A_a^H$ and $B_a^H$ are convolution integrals of 
splitting-like functions with different parton distribution functions.

To add fragmentation functions to the formalism we adopted the same idea 
as in the crossing case: we want to use the ``K'' factor already calculated.
Let us consider the process: $V \go H +(n-1)$~jets.  
First the NLO calculation for $V \go n$~jets is convoluted 
with all the appropriate fragmentation functions.  
Then, corrections must be applied in $n=4+\eps$ dimensions.
When the collinear contribution from a parton {\it h} splitting to parton
{\it i} and {\it j} is calculated for the ``K'' factor, the phase space 
is not only integrated over $S_{ij}$ up to $S_{min}$, but 
also over the momentum fraction of {\it i} compared to {\it h}.  
This last information is needed to properly add the fragmentation functions 
and should not be integrated over.  Furthermore, this contribution is 
convoluted with the parton {\it h} fragmentation function, instead of 
{\it i} or {\it j}.
The corrections for both of these effects are, along with
mass factorization, proportional to the Born cross section:
\beq
\als \, \sum_{h}\int dz \,
d\sigma_{h}^{LO}
\,  T_h^{H,scheme}(z) \ .
\eeq
The $T_a^H$ are the factorization scheme dependent ``tagging functions'', 
they have the same properties and functional $S_{min}$ dependence 
as the crossing functions.  
In the massive case, 
there is no collinear contribution included in the ``K'' factor of
the calculation without the fragmentation functions.
We derived the heavy quark tagging and crossing functions that 
implement the variable flavor number scheme~\cite{ACOT94}
(it defines the heavy quark factorization scheme).  This takes care of
large logarithms involving the heavy quark mass, leads to collinear
safe quantities in the sense 
that when the mass of the heavy quark tends to zero the massless 
result is recovered, and makes our formalism applicable
at any transverse energy.

\end{document}